\documentclass[11pt]{article}
\usepackage{hyperref}
\pdfoutput=1

\usepackage{hyperref}
\pdfoutput=1

\begin{document}
\title{How to freeze drops with powders}
\author{Jeremy O. Marston \\
\\Division of Physical Sciences and Engineering,
\\ King Abdullah University of Science and Technology,
\\ Thuwal, 23955-6900, Saudi Arabia}
\maketitle

\begin{abstract}
This document accompanies fluid dyanmics video entry V83911 for APS DFD 2012 meeting. In this video, we present experiments on how drop oscillations can be "frozen" using hydrophobic powders.
\end{abstract}

\section{Introduction}

In this video (ref), we show that when a liquid drop impacts a powder which is superhydrophobic, the drop rebounds with a powder coating that can "freeze" the oscillations and yield a deformed (i.e. non-spherical) liquid marble. For water drops, the critical impact speed for the onset of this phenomenon is $V_{c} = 1.6$ m/s. Repeat experiments with more viscous drops show even more deformed shapes.

For further information, please see the following article:
\begin{enumerate}
\item Marston et al. (2010) {\it Powder Technol.} {\bf202}, 223-236.
\item Marston et al. (2012) {\it Powder Technol.} {\bf228}, 424-428.
\end{enumerate}

\end{document}